\documentclass[
aps,%
11pt,%
final,%
notitlepage,%
oneside,%
twocolumn,%
nobibnotes,%
nofootinbib,%
superscriptaddress,%
noshowpacs,%
centertags]%
{revtex4}

\usepackage{multirow}
\renewcommand{\baselinestretch}{1.0}

\begin{document}

\selectlanguage{english}

\title{Halo Radius (Splashback Radius) of Groups and Clusters of Galaxies\\ on Small Scales}

\author{\firstname{F.~G.}~\surname{Kopylova}}
\email{flera@sao.ru}
\affiliation{\saoname}

\author{\firstname{A.~I.}~\surname{Kopylov}}
\email{akop@sao.ru}
\affiliation{\saoname}

\begin{abstract}
We report the results of a study of the distribution of galaxies in the projection along the radius (\mbox{$R \leq 3R_{200c}$})
for 157~groups and clusters of galaxies in the local Universe (0.01 < $z$ <
0.10) with line-of-sight velocity dispersions~200~km\,s$^{-1}$ < $ \sigma$ < 1100~km\,s$^{-1}$.
We introduce a new observed boundary for the halos of clusters of galaxies, which we identify with the splashback radius $R_{\rm{sp}}$.
We also identified the core of groups/clusters of galaxies with the radius  $R_c$. These radii are determined by the observed
integrated distribution of the number of galaxies as a function of squared angular radius from the center of the group/cluster,
which (usually) coincides with the brightest galaxy. We found for the entire sample that the boundary of dark matter  $R_{\rm{sp}}$
for groups/clusters of galaxies is proportional to the radius  $R_{\rm{200с}}$ of the virialized region. We measured the
mean radius $\langle R_{\rm{sp}} \rangle = 1.14\pm0.02$~Mpc for groups of galaxies ($\sigma \leq 400$~km\,s$^{-1}$) and
\mbox{$\langle R_{\rm{sp}} \rangle = 2.00\pm0.07$}~Mpc for clusters of galaxies ($\sigma > 400$~km\,s$^{-1}$).
The mean ratio of radii is \mbox{$\langle R_{\rm{sp}}/R_{\rm{200c}} \rangle = 1.40\pm0.02$}, or
$\langle R_{\rm{sp}}/R_{\rm{200m}} \rangle = 0.88\pm0.02$.


\end{abstract}

%\footnotetext{email: flera@sao.ru}
\maketitle

\section{INTRODUCTION}

Clusters of galaxies are the biggest gravitationally bound objects in the Universe. They are collapsed structures, which
are viewed as a dark-matter halo. Clusters of galaxies increase their mass as a result of impacts of galaxies
and smaller groups of galaxies along the filaments, and as a result of continuous infall of dark-matter objects.
They have no clear boundaries and their boundaries are defined based on the density contrast with respect to the critical or
mean density of the Universe. The evolution of clusters of galaxies is analyzed in terms of the spherical collapse model
in expanding Universe (Gunn and Gott 1972; Gott 1973). In this model the system of galaxies virializes in several oscillations:
maximum compression is followed by the next, secondary compression. Of special interest are the region of clusters of
galaxies where the model dark-matter particles (galaxies) from filaments occur  that fall for the first time onto the cluster
and the collapsing spherical halo. The  $N$-body simulations of the motion of the dark-matter halo (galaxies) performed by
Balogh et al. (2000); Mamon et al. (2004); Gill et al. (2005) revealed that a substantial fraction of such  particles (up to 50\%) located  outside virialized
regions of galaxy clusters  (out to 2$R{_{\rm vir}}$ or 2$R_{\rm{200c}}$\footnote{Here
$R_{\rm{200c}}$ (hereafter $R_{200}$) is the cluster radius inside which the density exceeds the critical density of the
Universe by a factor of 200. In our studies is determined by the dispersion of line-of-sight velocities of galaxies
in clusters. In simulations another radius---$R_{\rm{200m}}$---is often used, inside which the density exceeds the average
density of the Universe by a factor of 200.}) has already been inside the region
defined by  these boundaries. These are the so-called ``backsplash'' galaxies, which have once passed the pericenter of their orbit during
the gravitational collapse and which will return there again after reaching the apocenter.  Haines et al. (2015)
reports the results of Millennium simulations for 75~galaxy clusters at $z$ = 0.0 and shows that a substantial
fraction of \mbox{``backsplash'' galaxies} ``rebouns'' out to $3r_{\rm{proj}}/r_{200}$ on the phase diagram.
Pimbblet (2010) determined the
fraction of \mbox{``backsplash'' galaxies} inside 1--2\,$R{_{\rm vir}}$ based
on their analysis of the SDSS catalog and simulations of 14 galaxy clusters.

Adhikari et al. (2014) introduced the radius of a galaxy cluster  (the physical halo boundary),
the splashback radius $R_{\rm {sp}}$, as the radius where the recently accreted dark-matter  halos pile up
within the apocenters of their orbits. According to the results of the $N$-body simulations of
(Adhikari et al. 2014; Diemer and Kravtsov 2014) the $R_{\rm {sp}}$ radius shows up conspicuously by the sharp decrease on the
composite density profile of the dark-matter halo. The simulations performed by More et al. (2015) showed that
the location of $R_{\rm {sp}}$ depends on the mass accretion rate: \mbox{$R_{\rm {sp}} \sim 0.8$--$1R_{\rm {200m}}$}
in a cluster with high accretion rate, whereas \mbox{$R_{\rm{sp}}$} is of about  $1.5R_{\rm {200m}}$ in a cluster
with low accretion rate.  Fong and Hun (2021) determined the galaxy cluster radius, which contains the splashback orbits
of most of the galaxies, and which is equal to about 2--3$R_{\rm{sp}}$.

The authors of a number of studies search for and measure this radius in observed galaxy clusters (mostly distant ones
with $z > 0.1$). The $R_{\rm {sp}}$ radii were measured using various methods:
based on the surface density of galaxies (Adhikari et al. 2016; More et al. 2016; Baxter et al. 2017)
 or using the method of weak gravitational lensing (Umetsu and Diemer 2017; Chang et al 2018; Contigiani et al. 2019).
In their simulations, Bucsh and White (2017) used the same algorithm for identifying galaxy clusters as the one employed for real
observational data, and showed that the effect of the projection of background galaxies onto the cluster may
distort the measured  $R_{\rm {sp}}$ radius. A sample of galaxy clusters selected based not on the optical parameters
but rather of the SZ signal, which better correlates with mass, also made it possible to measure the $R_{\rm {sp}}$
radii (Shin et al. 2019; Z\'{u}rcher and More 2019). Measuring the $R_{\rm {sp}}$ radius (the size of the dark-matter halo) in galaxy clusters
using various methods is of great importance for the study of the large-scale structure of the Universe.

In this study we analyze observational manifestations of the splashback feature in a sample of groups and clusters of galaxies
(based on SDSS data). Earlier, we found the  ``boundary'' of 29 galaxy clusters clearly identified by the integrated distribution
of the squared clustercentric distance  of all cluster galaxies. We call this  boundary the halo radius $R_h$ (Kopylov and Kopylova 2015).
This radius is usually greater than $R_{\rm{200}}$
and is measured from the projected profile,
i.e., by the point of transition from the steep increase of the number of galaxies at the cluster center to its linear decrease.
We later identified this radius with the splashback radius $R_{\rm{sp}}$ reported the results for about  100~groups and
clusters of galaxies (Kopylov and Kopylova 2015; Kopylova and Kopylov 2016, 2018, 2019).
Kopylova and Kopylov (2018, 2019) showed that the distribution of early-type galaxies in clusters can be used to more
precisely estimate the required radius.
We analyzed the data for 40~systems to find the mean radius to be \mbox{$R_{\rm{sp}} = 1.54\pm0.06~R_{200}$}
or \mbox{$R_{\rm{sp}} = 0.96\pm0.06~R_{\rm{200m}}$}
(given that  \mbox{4$R_{200} \approx 2.5R_{\rm{200m}}$)},
which varies from  1.10~Mpc for the NGC\,5627 group with \linebreak $\sigma$ = 314~km~$s^{-1}$ to 4.17~Mpc for the Coma
cluster  (A\,1656) with \mbox{$\sigma$ = 921~km\,$s^{-1}$}.

In this study we increased the size of our sample to  157~groups and clusters of galaxies from the regions
of  Leo, Hercules, Ursa Major, Corona Borealis, and Bootes superclusters, from the region of  A\,1656/\,A1367,
supercluster, as well as smaller superclusters and the field.

In our study we used the data from the SDSS (Sloan Digital Sky Survey, DRs 7, 8) and
2MASS XSC (2MASX, Two-Micron ALL-Sky Survey Extended Source Catalog) catalogs and from
NED (NASA Extragalactic Database).

The paper has the following layout. Section~2 describes the sample of groups/clusters of galaxies and the
procedure we use to determine the radii $R_{\rm{sp}}$ and $R_c$ of the objects studied. Section~3
presents the construction of the dependences of the inferred radii on the basic parameters of groups and
clusters of galaxies, which we determined in earlier studies.  Section~4 summarizes the main results of this work.
We use the following values of cosmological parameters: $\Omega_m=0.3$, $\Omega_{\Lambda}=0.7$,
$H_0=70$~km~s$^{-1}$ Mpc$^{-1}$.

\section{DESCRIPTION OF THE DATA AND THE TECHNIQUUE FOR MEASURING THE RADII}

Our sample consists of  157~groups and clusters from the regions of  Leo ($N=12$), Hercules (\mbox{$N=27$}),
Ursa Major (\mbox{$N=19$}), Corona Borealis ($N=7$), Bootes
\mbox{($N=13$)}, and other smaller superclusters and field regions ($N=11$,
$N=20$), and groups of galaxies from the region of A\,1656/\,A1367 supercluster \mbox{($N=48$)}.
The sample of galaxies in superclusters was compiled for measuring the peculiar velocities of their constituent
galaxy clusters. In addition, we also analyzed the dependence between the dynamical mass within the virialized radius
$R_{200}$ and the $K$-band infrared luminosity (based on  2MASX data).
X-ray radiation was recorded from all galaxy clusters except 21 groups of galaxies. We adopt their X-ray luminosities
from catalogs of galaxy clusters based on ROSAT data  (see the above papers for more detailed references).
Our sample of groups and clusters of galaxies spans the maximum interval of line-of-sight velocities ranging from  200
to 1100~km~s$^{-1}$ and have redshifts  $ 0.01 < z < 0.10$.

For these groups and clusters
we measured the heliocentric redshifts, line-of-sight velocity dispersions with cosmological correction
$(1+z)^{-1}$ applied, the $R_{200}$ radii, $K$-band luminosities $L_{K,200}$ (\mbox{$M_K
<-21\,.\!\!^{\rm m}0$}), dynamical masses $M_{200}$, and other parameters within the  $R_{200}$ radius (Kopylov and Kopylova 2015;
Kopylova and Kopylov 2016, 2017, 2018, 2019). The physical properties of the galaxy groups and clusters studied are listed in \linebreak \mbox{Columns~(1)--(13)}
of Table~\ref{data1}.

The empirical cluster radius $R_{200}$
can be predicted based on the dispersion of line-of-sight velocities provided that the cluster mass satisfies
the condition $M(r)\propto r$. This radius can be estimated by the formula
$R_{200} = \sqrt{3}\,\sigma /(10H(z))$\,Mpc (Carlberg et al. 1997).
In this case, if we assume that the cluster is virialized inside this radius, its mass can be found by the formula
$M_{200} =
3\,G^{-1}R_{200}\,\sigma^{2}$, where $\sigma$ is the one-dimensional dispersion of the line-of-sight velocities of galaxies
located inside the $R_{200}$ radius and $G$ is the gravitational constant. Thus the cluster mass that we measure  $M_{200} \propto
\sigma^3$. The mass $M_{200}$ contained in the  spherical halo of radius $R_{200}$ can also be measured
directly based on the critical density, which depends on  $z$: $M_{200} =
\dfrac{4}{3} \pi R_{200}^{3} \times 200\rho_c$.

\onecolumngrid
\setcaptionwidth{\linewidth}
\setcaptionmargin{0mm} %
\onelinecaptionstrue \captionstyle{normal}
\medskip
\renewcommand{\baselinestretch}{1.0}
\begin{longtable}{l|c|c|c|c|c|c|c|c|c|c|c|c}
   \caption{Dynamical parameters of groups and clusters of galaxies:
(1)---the name of the cluster of galaxies;
(2)---the heliocentric redshift; (3)---the line-of-sight velocity dispersion with cosmological correction
$(1+z)^{-1}$ applied; (4)---the halo mass $\log M_{200}/M_{\odot}$;
(5)---the $K$ -- luminosity $\log L_{K,200}/L_{\odot}$; (6)---the X-ray (\mbox{0.1--2.4}~keV) luminosity
$\log L_X$ from X-ray catalogs of galaxy clusters based on  ROSAT observations
(we recomputed the luminosities using our inferred line-of-sight velocities of clusters and the adopted model);
(7)---references to the X-ray luminosity;
(8)---the  $R_{200}$ radius in Mpc; (9)---the $R_{\rm{sp}}$ radius in Mpc; (10)---the $R_c$ radius in Mpc;
(11)---$R_{\rm{sp}}/R_{200}$;
(12)---the magnitude gap  $\Delta M_{1,4}$ between the first and fourth brightest galaxies, and
(13)---the concentration $\Sigma_5$ of galaxies brighter than $M_K=-23\,.\!\!^{\rm m}3$ calculated by using the projected distance to the fifth galaxy closest to the center.\\
{\small
$^a$ X-ray references: (1) Kopylova and Kopylov (2009),
(2) Kopylova and Kopylov (2011), (3) Kopylova and Kopylov (2013), (4) Bo\"{o}hringer et al. (2000),
(5) Ebeling et al.(1998), (6) Kopylova and Kopylov (2016), (7) Kopylov and Kopylova (2007), (8)
Kopylov and Kopylova (2012), (9) Kopylov and Kopylova (2010), (10) Kopylova and Kopylov (2017), (11) Bo\"{o}hringer et al. (2004),
(12) Mahdavi et al. (2000), (13) Ebeling et al. (2002), (14) Mulchaey et al. (2003), (15) Ledlow et al. (2003).
}
}\label{data1}\\
    \hline
\multicolumn{1}{c|}{\multirow{2}{*}{System}} & \multicolumn{1}{c|}{\multirow{2}{*} {$z_h$}} & $\sigma$,   & $\log M_{200}$, & $\log L_{K,\,200}$, & $\log L_X$, & \multicolumn{1}{c|}{\multirow{2}{*} {Ref$^{a}$}}  & $R_{200}$, & $R_{\rm{sp}}$, & $R_c$, & \multicolumn{1}{c|}{\multirow{2}{*}{$R_{\rm{sp}}/R_{200}$}} & \multicolumn{1}{c|}{\multirow{2}{*} {$\Delta M_{1,4}$}} & $\Sigma_5$,  \\
&          & km~s$^{-1}$& [$M_{\odot}$] &  [$L_{\odot}$] & \,erg\,s$^{-1}$ &   & Mpc       &  Mpc     &  Mpc  &    &  & Mpc$^{-2}$               \\
    \hline
\multicolumn{1}{c|}{(1)}&  (2) & (3)   & (4) & (5)& (6)  &(7) &(8)&(9)& (10)&(11)& (12) & (13) \\
  \hline
    \endfirsthead
    \caption{(Continued)}\\
    \hline
\multicolumn{1}{c|}{\multirow{2}{*}{System}} & \multicolumn{1}{c|}{\multirow{2}{*} {$z_h$}} & $\sigma$,   & $\log M_{200}$, & $\log L_{K,\,200}$, & $\log L_X$, & \multicolumn{1}{c|}{\multirow{2}{*} {Ref$^a$}} & $R_{200}$, & $R_{\rm{sp}}$, & $R_c$, & \multicolumn{1}{c|}{\multirow{2}{*}{$R_{\rm{sp}}/R_{200}$}}& \multicolumn{1}{c|}{\multirow{2}{*} {$\Delta M_{1,4}$}} &$\Sigma_5$,  \\
&          & km~s$^{-1}$& [$M_{\odot}$] &  [$L_{\odot}$] & erg\,s$^{-1}$&  & Mpc       &  Mpc     &  Mpc  &    &  & Mpc$^{-2}$                    \\
\hline
\multicolumn{1}{c|}{(1)}&  (2) & (3)   & (4) & (5)& (6)  &(7) &(8)&(9)& (10)&(11)& (12) & (13) \\
   \hline
    \endhead
    \hline
    \endfoot
    \hline
    \endlastfoot

HCG\,42   &   0.012588&  228& 13.30& 11.79& 42.28&(6)&  0.56& 0.75& 0.30&   1.34& 2.31& 0.54\\
AWM\,3    &   0.014878&  269& 13.52& 11.68&<42.00&(12)&  0.66& 1.04& 0.44&   1.58& 1.55& 0.52\\
NGC\,2563 &   0.015701&  369& 13.93& 12.15& 42.23&(6)&  0.91& 1.00& 0.39&   1.10& 1.30& 1.23\\
AWM\,7    &   0.017344&  698& 14.77& 12.62& 43.98&(13)   &  1.71& 1.97& 1.08&   1.15& 2.28& 2.07\\
NGC\,0533 &   0.018411&  404& 14.05& 12.25& 42.84&(6)&  0.99& 1.40& 0.57&   1.41& 3.24& 0.32\\
NGC\,0741 &   0.018416&  368& 13.93& 12.10& 42.60&(6)&  0.90& 0.98& 0.39&   1.09& 2.98& 0.66\\
NGC\,0080 &   0.019098&  296& 13.64& 12.29& 42.84&(6)&  0.73& 0.79& 0.37&   1.08& 1.31& 1.48\\
MKW\,04   &   0.020208&  515& 14.37& 12.41& 43.15&(11)   &  1.26& 1.53& 0.78&   1.21& 1.84& 1.72\\
NGC\,3022 &   0.020959&  276& 13.56& 11.97&<42.30&(12)&  0.68& 1.14& 0.50&   1.68& 1.50& 1.20\\
A\,1367   &   0.021743&  749& 14.86& 12.89& 43.91&(5)&  1.84& 2.74& 1.38&   1.49& 0.89& 2.25\\
NGC\,2783 &   0.022151&  346& 13.85& 12.04& 42.00&(14)   &  0.85& 1.02& 0.59&   1.20& 3.04& 0.60\\
UGC\,07115&   0.022199&  334& 13.81& 12.12& 42.60&(6)&  0.82& 1.06& 0.66&   1.29& 1.79& 0.78\\
UGC\,02005&   0.022342&  352& 13.88& 12.05& 42.60&(6)&  0.86& 1.12& 0.59&   1.30& 1.43& 0.72\\
IC\,5357  &   0.022436&  381& 13.98& 12.08& 42.70&(6)&  0.93& 1.02& 0.57&   1.10& 0.96& 1.00\\
NGC\,1016 &   0.022581&  322& 13.76& 12.28&<42.30&(12)   &  0.79& 1.10& 0.40&   1.39& 2.01& 1.11\\
NGC\,3158 &   0.022630&  375& 13.95& 12.25& 42.48&(6)&  0.92& 1.24& 0.32&   1.35& 2.01& 1.67\\
NGC\,0070 &   0.022645&  415& 14.09& 12.30& 42.95&(6)&  1.02& 1.15& 0.62&   1.13& 0.31& 2.72\\
AWM\,2    &   0.022761&  293& 13.63& 11.97&  --  &   &  0.72& 0.78& 0.42&   1.08& 2.18& 0.64\\
NGC\,5171 &   0.023000&  371& 13.94& 12.18& 43.00&(6)&  0.91& 1.47& 0.58&   1.62& 0.84& 1.63\\
NGC\,2832 &   0.023044&  331& 13.79& 12.25& 43.00&(6)&  0.81& 1.25& 0.63&   1.54& 2.34& 1.32\\
A\,1656   &   0.023250&  921& 15.13& 13.22& 44.57&(5)&  2.26& 4.17& 2.00&   1.84& 1.78& 2.02\\
NGC\,5129 &   0.023402&  290& 13.62& 12.11& 42.95&(6)&  0.71& 1.36& 0.55&   1.92& 3.01& 0.84\\
MCG-0129015&  0.023813&  334& 13.81& 12.02& 42.64& (12)  &  0.82& 1.56& 0.60&   1.90& 1.72& 0.66\\
NGC\,7436B&   0.024720&  383& 13.98& 12.24&<42.00&(12)   &  0.94& 1.08& 0.48&   1.15& 1.84& 1.53\\
NGC\,5306 &   0.024732&  305& 13.68& 12.06& 42.70&(6)&  0.75& 1.30& 0.58&   1.73& 3.22& 0.78\\
NGC\,5223 &   0.024834&  271& 13.53& 12.19& 42.78&(6)&  0.66& 0.97& 0.36&   1.47& 2.09& 1.20\\
MKW\,05   &   0.024858&  288& 13.61& 11.84&  --  &   &  0.70& 0.88& 0.26&   1.26& 2.12& 0.46\\
NGC\,4325 &   0.025386&  271& 13.53& 11.79& 42.71&(6)&  0.66& 0.84& 0.41&   1.27& 2.04& 0.40\\
IC\,01867 &   0.026023&  318& 13.74& 12.10&<42.30&(6)&  0.78& 1.16& 0.58&   1.49& 1.97& 0.93\\
NGC\,7237 &   0.026102&  376& 13.96& 12.20& 42.75&(6)&  0.92& 1.58& 0.60&   1.72& 1.26& 1.45\\
IC\,02476  &   0.026198&  243& 13.38& 11.88&<42.30&(12)   &  0.59& 0.96& 0.32&   1.63& 2.12& 0.71\\
NGC\,5627 &   0.026682&  314& 13.72& 12.14& 42.30&(6)&  0.77& 1.10& 0.53&   1.43& 2.49& 0.85\\
MKW\,08   &   0.026906&  450& 14.19& 12.50& 43.48&(3)&  1.10& 1.73& 0.85&   1.57& 0.63& 2.00\\
UGC\,05088&   0.027622&  247& 13.41& 11.72& 42.30&(3)&  0.60& 0.90& 0.41&   1.50& 1.79& 0.43\\
MKW\,04s  &   0.027928&  423& 14.11& 12.27& 43.04&(5)&  1.03& 1.53& 0.78&   1.48& 2.17& 1.28\\
AWM\,1    &   0.028652&  402& 14.05& 12.37&<42.30&(12)   &  0.98& 1.15& 0.62&   1.17& 1.14& 1.61\\
NGC\,2795 &   0.028992&  431& 14.14& 12.38& 42.70&(3)&  1.04& 1.32& 0.62&   1.27& 1.94& 1.00\\
NGC\,6338 &   0.029342&  552& 14.46& 12.45& 43.40&(3)&  1.35& 2.12& 0.50&   1.57& 2.00& 1.65\\
NGC\,3119 &   0.029657&  355& 13.88& 12.19& 42.60&(6)&  0.87& 1.20& 0.60&   1.38& 2.50& 1.61\\
NGC\,5758 &   0.029923&  291& 13.62& 12.09& 42.84&(6)&  0.71& 1.09& 0.50&   1.54& 1.11& 1.49\\
A\,2199   &   0.030458&  746& 14.85& 13.01& 44.31&(3)&  1.82& 3.56& 1.42&   1.96& 0.70& 2.36\\
A\,2197   &   0.030477&  547& 14.45& 12.85& 43.08&(3)&  1.34& 1.80& 0.92&   1.34& 1.54&  --  \\
NGC\,6107 &   0.031093&  546& 14.44& 12.55& 43.23&(3)&  1.33& 1.90& 1.03&   1.43& 0.45& 1.32\\
NGC\,6159 &   0.031320&  266& 13.51& 11.81& 42.78&(3)&  0.65& 0.89& 0.45&   1.37& 2.60& 0.48\\
AWM\,4    &   0.031827&  380& 13.97& 12.04& 43.36&(3)&  0.93& 1.34& 0.57&   1.44& 3.13& 0.61\\
A\,0999   &   0.031866&  248& 13.41& 12.05& 42.48&(2)&  0.60& 1.11& 0.53&   1.85& 2.18& 1.18\\
UGC\,04991&   0.031958&  515& 14.37& 12.37& 42.60&(3)&  1.26& 1.72& 0.65&   1.36& 1.25& 1.73\\
A\,2162   &   0.032147&  346& 13.85& 12.17& 42.60&(3)&  0.84& 1.12& 0.63&   1.33& 1.89& 1.08\\
A\,1177   &   0.032159&  337& 13.81& 12.09& 43.04&(2)&  0.82& 1.14& 0.55&   1.39& 2.33& 0.85\\
A\,1016   &   0.032178&  267& 13.51& 12.09&  --  &   &  0.65& 1.02& 0.35&   1.57& 1.76& 1.11\\
A\,1314   &   0.032443&  494& 14.31& 12.49& 43.11&(2)&  1.18& 1.90& 0.55&   1.61& 0.97& 1.46\\
A\,1185   &   0.032734&  676& 14.72& 12.84& 43.18&(2)&  1.69& 2.19& 0.84&   1.30& 1.06& 2.28\\
A\,1257   &   0.034588&  242& 13.38& 11.91&  --  &   &  0.58& 0.99& 0.40&   1.71& 1.08& 1.28\\
A\,2063   &   0.034664&  753& 14.86& 12.79& 44.01&(3)&  1.83& 2.61& 1.10&   1.43& 0.90& 2.00\\
A\,2052   &   0.034726&  623& 14.61& 12.70& 44.11&(3)&  1.52& 2.12& 0.87&   1.39& 1.28& 1.80\\
AWM\,5    &   0.035043&  517& 14.37& 12.67& 43.45&(3)&  1.24& 1.70& 0.89&   1.37& 1.93& 1.00\\
A\,1228A  &   0.035055&  216& 13.23& 12.18&  --  &   &  0.57& 1.00& 0.47&   1.75& 0.73& 1.36\\
RXC\,J1057.7+3739&  0.035208&  297& 13.65& 11.93& 42.48&(2)&  0.72& 0.97& 0.46&   1.35& 1.69& 1.08\\
VV\,196   &   0.035289&  412& 14.08& 12.12& 42.85&(3)&  1.00& 1.17& 0.71&   1.17& 1.09& 2.48\\
A\,2147   &   0.036179&  853& 15.02& 13.11& 44.20&(3)&  2.08& 3.47& 1.49&   1.67& 0.56& 1.82\\
A\,2151   &   0.036378&  734& 14.83& 13.08& 43.65&(3)&  1.79& 2.10& 0.55&   1.17& 0.63& 1.68\\
NGC\,5098 &   0.036812&  445& 14.18& 12.46& 43.11&(3)&  1.08& 1.73& 0.71&   1.60& 0.92& 1.93\\
RXC\,J1511.5+0145&  0.038990&  374& 13.95& 12.10& 42.95&(3)&  0.91& 1.12& 0.63&   1.23& 1.65& 0.80\\
A\,1139   &   0.039327&  459& 14.21& 12.57& 43.18&(2)&  1.12& 1.64& 0.71&   1.46& 0.84& 1.79\\
RBS\,858  &   0.039586&  445& 14.18& 12.39& 43.11&(2)&  1.08& 1.64& 0.84&   1.52& 1.81& 1.04\\
A\,2107   &   0.041335&  581& 14.52& 12.69& 43.77&(3)&  1.41& 2.17& 0.63&   1.54& 1.40& 1.61\\
A\,1228B  &   0.042892&  347& 13.85& 12.25& 42.78&(2)&  0.84& 1.28& 0.50&   1.52& 2.70& 0.70\\
A\,1983   &   0.044803&  460& 14.22& 12.70& 43.41&(3)&  1.12& 1.34& 0.78&   1.20& 0.62& 2.24\\
MKW\,03s  &   0.044953&  608& 14.58& 12.67& 44.15&(3)&  1.47& 1.95& 0.76&   1.33& 0.66& 1.74\\
A\,0957   &   0.045026&  689& 14.74& 12.70& 43.61&(3)&  1.67& 1.79& 0.99&   1.07& 2.23& 1.93\\
A\,2040   &   0.045242&  589& 14.54& 12.82& 43.26&(3)&  1.43& 2.20& 0.84&   1.54& 0.58& 2.62\\
RXC\,J1010.3+5430&  0.045877&  384& 13.98& 12.36& 42.30&(1)&  0.93& 1.20& 0.81&   1.29& 1.92& 0.73\\
A\,1100   &   0.046463&  402& 14.04& 12.31& 42.78&(3)&  0.97& 1.22& 0.55&   1.26& 1.74& 1.77\\
RXC\,J1722.2+3042&  0.046580&  524& 14.39& 12.56& 42.70&(3)&  1.27& 1.64& 0.67&   1.29& 1.78& 1.40\\
RXC\,J0748.2+1833&  0.046602&  454& 14.20& 12.60& 43.15&(3)&  1.09& 1.28& 0.71&   1.17& 1.73& 2.24\\
SHK\,352  &   0.049521&  532& 14.41& 12.60& 43.43&(4)&  1.29& 2.17& 0.74&   1.68& 0.64& 2.22\\
A\,0671   &   0.049802&  805& 14.95& 12.89& 43.66&(6)&  1.95& 1.97& 0.84&   1.01& 1.68& 2.26\\
Sh\,166   &   0.050043&  323& 13.76& 12.17&  --  &   &  0.78& 1.38& 0.56&   1.77& 1.22& 1.52\\
Z\,2844  &    0.050489&  401& 14.04& 12.39& 43.46&(5)&  0.97& 1.48& 0.63&   1.52& 2.05& 1.30\\
A\,0757   &   0.051319&  368& 13.92& 12.43& 43.66&(6)&  0.89& 1.79& 0.77&   2.01& 0.46& 1.85\\
A\,1291A  &   0.051349&  391& 14.00& 12.17& 43.34&(1)&  0.94& 1.48& 0.55&   1.57& 1.32& 0.86\\
A\,1377   &   0.051807&  632& 14.63& 12.77& 43.45&(1)&  1.53& 2.05& 0.63&   1.34& 0.71& 2.27\\
A\,1461   &   0.053962&  317& 13.73& 11.92&   -- &   &  0.76& 1.13& 0.49&   1.49& 1.22& 1.15\\
RXC\,J1022.2+3831&  0.054163&  551& 14.45& 12.65& 43.26&(6)&  1.33& 1.57& 0.84&   1.18& 0.71& 1.45\\
RXC\,J0844.9+4258&  0.054858&  320& 13.74& 12.16& 42.90&(10)&  0.77& 1.30& 0.63&   1.69& 2.25& 0.86\\
RXC\,J1122.2+6713&  0.055119&  237& 13.34& 11.79& 42.78&(1)&  0.57& 1.07& 0.42&   1.88& 2.42& 0.68\\
A\,1318   &   0.056419&  394& 14.01& 12.48& 42.60&(1)&  0.95& 1.26& 0.68&   1.33& 1.19& 1.40\\
RXC\,J1654.4+2334&  0.057075&  383& 13.98& 12.27& 43.26&(10)&  0.92& 1.26& 0.77&   1.37& 1.70& 1.30\\
A\,1291B  &   0.057161&  396& 14.02& 12.07&  --  &   &  0.95& 1.05& 0.50&   1.11& 1.34& 1.04\\
A\,2169   &   0.057656&  502& 14.33& 12.46& 43.36&(5)&  1.21& 1.45& 0.63&   1.20& 0.44& 1.26\\
A\,1991   &   0.058463&  554& 14.46& 12.81& 43.85&(6)&  1.33& 1.49& 0.95&   1.12& 0.99& 1.59\\
RXC\,J0746.7+3059&  0.058482&  317& 13.73& 12.21& 43.20&(4)&  0.76& 0.90& 0.56&   1.18& 0.63& 2.20\\
A\,1383   &   0.059583&  464& 14.23& 12.58& 43.11&(1)&  1.12& 1.67& 0.95&   1.49& 0.90& 1.68\\
A\,1507   &   0.059967&  432& 14.13& 12.42& 42.85&(1)&  1.02& 1.53& 0.63&   1.50& 1.19& 1.08\\
A\,0602   &   0.060551&  560& 14.47& 12.64& 43.76&(6)&  1.35& 1.90& 0.92&   1.41& 0.31& 1.51\\
RXC\,J1224.8+3156&  0.060664&  454& 14.20& 12.40& 43.18&(4)&  1.09& 1.73& 0.95&   1.59& 0.70& 1.52\\
Anon\,4   &   0.061053&  397& 14.02& 12.39&  --  &   &  0.96& 1.52& 0.64&   1.58& 1.47& 1.64\\
A\,1452   &   0.061649&  408& 14.06& 12.28&  --  &   &  0.98& 1.10& 0.59&   1.12& 1.21& 1.28\\
A\,1781   &   0.062264&  362& 13.90& 12.44&  --  &   &  0.87& 1.56& 0.65&   1.79& 1.35& 1.28\\
A\,1795   &   0.062444&  775& 14.89& 12.95& 44.76&(6)&  1.86& 3.06& 1.34&   1.64& 1.39& 2.09\\
A\,1275   &   0.062750&  348& 13.85& 12.29& 43.34&(6)&  0.84& 1.20& 0.67&   1.43& 1.58& 1.11\\
A\,1003   &   0.062763&  575& 14.50& 12.49& 43.00&(1)&  1.38& 1.66& 0.77&   1.20& 1.32& 1.36\\
RXC\,J1351.7+4622&  0.062915&  528& 14.40& 12.60& 43.08&(4)&  1.29& 1.70& 0.92&   1.32& 2.01& 0.90\\
A\,1831A  &   0.062942&  480& 14.27& 12.48&  --  &   &  1.15& 1.43& 0.69&   1.24& 1.12& 1.57\\
A\,1825   &   0.063274&  633& 14.63& 12.56& 43.04&(6)&  1.52& 1.38& 0.55&   0.91& 0.95& 1.70\\
A\,1668   &   0.063699&  635& 14.63& 12.72& 43.91&(5)&  1.52& 1.82& 0.95&   1.20& 1.42& 1.85\\
A\,1436   &   0.064960&  700& 14.76& 12.83& 43.72&(1)&  1.68& 1.82& 1.41&   1.08& 0.37& 1.76\\
A\,2149   &   0.065253&  361& 13.90& 12.46& 43.62&(10)&  0.87& 1.28& 0.63&   1.47& 2.23& 1.57\\
A\,1775A  &   0.065591&  324& 13.76& 12.31&  --  &   &  0.78& 1.64& 0.45&   2.10& 1.07& 1.56\\
A\,2124   &   0.065722&  736& 14.83& 12.83& 43.84&(6)&  1.77& 2.21& 0.92&   1.25& 1.73& 1.84\\
A\,2079   &   0.065746&  618& 14.60& 12.95& 43.57&(6)&  1.48& 2.12& 0.95&   1.43& 0.97& 1.26\\
RXC\,J1206.6+2215&  0.065786&  269& 13.52& 12.08& 43.15&(10)&  0.65& 1.22& 0.60&   1.88& 2.41& 1.18\\
A\,2092   &   0.066564&  486& 14.29& 12.52& 43.60&(6)&  1.17& 2.05& 0.76&   1.75& 0.80& 2.17\\
Anon\,3   &   0.067960&  380& 13.96& 12.29&  --  &   &  0.91& 1.26& 0.50&   1.38& 1.25& 1.52\\
A\,1035A  &   0.067997&  563& 14.48& 12.63& 42.85&(7)&  1.35& 1.79& 0.90&   1.33& 0.35& 1.72\\
A\,1569A  &   0.068759&  484& 14.26& 12.44& 43.30&(8)&  1.16& 1.50& 0.60&   1.29& 0.82& 2.82\\
A\,1371   &   0.068891&  552& 14.45& 12.67& 43.53&(4)&  1.32& 2.12& 0.90&   1.61& 0.86& 1.91\\
A\,1066   &   0.068917&  768& 14.88& 12.98& 43.82&(4)&  1.84& 3.03& 1.26&   1.65& 0.95& 1.63\\
A\,1270   &   0.068939&  524& 14.38& 12.68& 42.78&(1)&  1.26& 1.41& 0.95&   1.12& 0.34& 1.92\\
A\,1534   &   0.069848&  322& 13.75& 12.41&  --  &   &  0.77& 1.40& 0.57&   1.82& 1.96& 1.26\\
Anon\,1   &   0.069884&  608& 14.58& 12.70& 43.54&(1)&  1.47& 1.76& 1.14&   1.20& 1.21& 1.79\\
A\,1767   &   0.070326&  816& 14.96& 13.03& 44.10&(5)&  1.95& 2.55& 1.10&   1.31& 1.50& 2.05\\
Z\,6718   &   0.071374&  550& 14.45& 12.54& 43.80&(5)&  1.32& 2.00& 0.87&   1.52& 1.04& 1.81\\
A\,1904   &   0.071708&  771& 14.89& 13.07& 43.67&(15)   &  1.84& 2.02& 1.12&   1.10& 1.10& 2.21\\
RXC\,J1054.2+5450A&  0.071886&  507& 14.34& 12.68& 43.72&(1)&  1.21& 1.52& 0.93&   1.26& 0.29& 2.05\\
A\,1589   &   0.071955&  778& 14.90& 13.06& 44.23&(8)&  1.86& 3.11& 1.55&   1.67& 0.96& 1.74\\
A\,2065   &   0.072211& 1104& 15.35& 13.31& 44.40&(6)&  2.64& 3.36& 1.67&   1.27& 0.54& 2.82\\
A\,1024   &   0.073296&  578& 14.51& 12.60& 43.40&(4)&  1.38& 1.83& 0.59&   1.33& 1.43& 2.58\\
A\,2089   &   0.073546&  531& 14.40& 12.66& 43.15&(15)   &  1.27& 1.48& 0.91&   1.16& 1.44& 1.30\\
A\,2064   &   0.073689&  633& 14.63& 12.66& 43.86&(4)&  1.51& 1.67& 1.00&   1.11& 1.96& 1.79\\
J1051.8+5523B  &   0.073762&  420& 14.09& 12.49& --& &  1.00& 1.22& 0.71&   1.22& 1.75& 1.53\\
A\,1238   &   0.074111&  541& 14.42& 12.78& 43.41&(6)&  1.29& 2.12& 0.90&   1.64& 1.03& 2.66\\
A\,1775B  &   0.075138&  581& 14.52& 12.78& 44.20&(6)&  1.39& 1.78& 0.81&   1.28& 1.95& 1.65\\
A\,1203   &   0.075307&  416& 14.08& 12.64&  --  &   &  0.99& 1.45& 0.67&   1.46& 1.32& 1.41\\
A\,1800   &   0.075321&  705& 14.77& 12.91& 44.19&(6)&  1.68& 2.28& 1.24&   1.36& 2.02& 1.86\\
A\,1190   &   0.075334&  670& 14.70& 12.91& 43.88&(5)&  1.60& 1.95& 0.89&   1.22& 1.12& 1.64\\
A\,1831B  &   0.075481&  952& 15.16& 12.97& 44.18&(9)&  2.27& 2.85& 1.64&   1.26& 1.17& 2.19\\
A\,1424   &   0.075900&  632& 14.63& 12.82& 43.71&(4)&  1.51& 1.82& 1.12&   1.20& 1.33& 2.02\\
A\,1205   &   0.076103&  787& 14.91& 12.98& 44.02&(4)&  1.88& 2.26& 1.34&   1.20& 1.76& 1.74\\
A\,1516   &   0.076166&  660& 14.68& 12.87&  --  &   &  1.58& 1.82& 1.34&   1.15& 1.53& 2.23\\
A\,1173   &   0.076193&  516& 14.36& 12.61& 43.70&(5)&  1.23& 1.97& 0.67&   1.60& 1.21& 1.89\\
J1350.2+2913 &   0.076439&  359& 13.89& 12.40& --&   &  0.86& 1.30& 0.41&   1.51& 0.89& 1.83\\
Z\,4905  &    0.076817&  568& 14.49& 12.67& 43.79&(5)&  1.36& 2.00& 1.00&   1.47& 1.73& 1.73\\
Z\,5029  &    0.077360&  912& 15.10& 13.15& 44.43&(5)&  2.18& 2.55& 1.48&   1.17& 0.92& 2.58\\
A\,1773   &   0.077425&  832& 14.98& 12.98& 43.89&(5)&  1.98& 2.43& 1.10&   1.23& 1.20& 2.54\\
A\,2061   &   0.077746&  712& 14.78& 13.12& 44.31&(6)&  1.70& 1.95& 0.97&   1.15& 1.15& 1.88\\
A\,2029   &   0.077812& 1046& 15.28& 13.37& 44.89&(6)&  2.50& 4.24& 1.53&   1.70& 2.35& 2.21\\
A\,1780   &   0.077855&  474& 14.25& 12.70&  --  &   &  1.13& 1.82& 0.78&   1.61& 0.90& 1.76\\
A\,1035B  &   0.078276&  613& 14.59& 12.58& 43.30&(7)&  1.46& 1.61& 0.84&   1.10& 1.47& 2.07\\
A\,1898   &   0.078525&  434& 14.13& 12.47& 43.04&(6)&  1.04& 1.90& 0.81&   1.83& 0.50&  --\\
A\,1809   &   0.079290&  729& 14.81& 12.99& 43.91&(5)&  1.74& 2.47& 1.10&   1.42& 1.45& 1.65\\
A\,1569B  &   0.079331&  493& 14.30& 12.46&  --  &   &  1.18& 1.69& 0.67&   1.43& 1.71& 1.88\\
A\,2019   &   0.081176&  345& 13.83& 12.27& 43.18&(6)&  0.82& 1.22& 0.59&   1.49& 1.78& 1.34\\
A\,1750   &   0.085934&  747& 14.84& 13.15& 44.32&(11)&  1.78& 2.30& 1.23&   1.29& 1.24& 2.00\\
A\,2245   &   0.087950& 1037& 15.27& 13.25& 43.67&(5)&  2.46& 2.93& 1.10&   1.19& 1.45& 1.99\\
A\,2142   &   0.090135&  963& 15.17& 13.42& 45.02&(6)&  2.28& 4.12& 1.52&   1.81& 0.83& 1.83\\
A\,2244   &   0.098993& 1049& 15.28& 13.18& 44.68&(5)&  2.48& 3.19& 1.48&   1.28& 1.85& 1.87\\

        \hline
\end{longtable} \twocolumngrid

\renewcommand{\baselinestretch}{1.0}

\begin{figure*}[]
    \setcaptionmargin{5mm} \onelinecaptionstrue \captionstyle{normal}
    \includegraphics[scale=0.53, angle=-90]{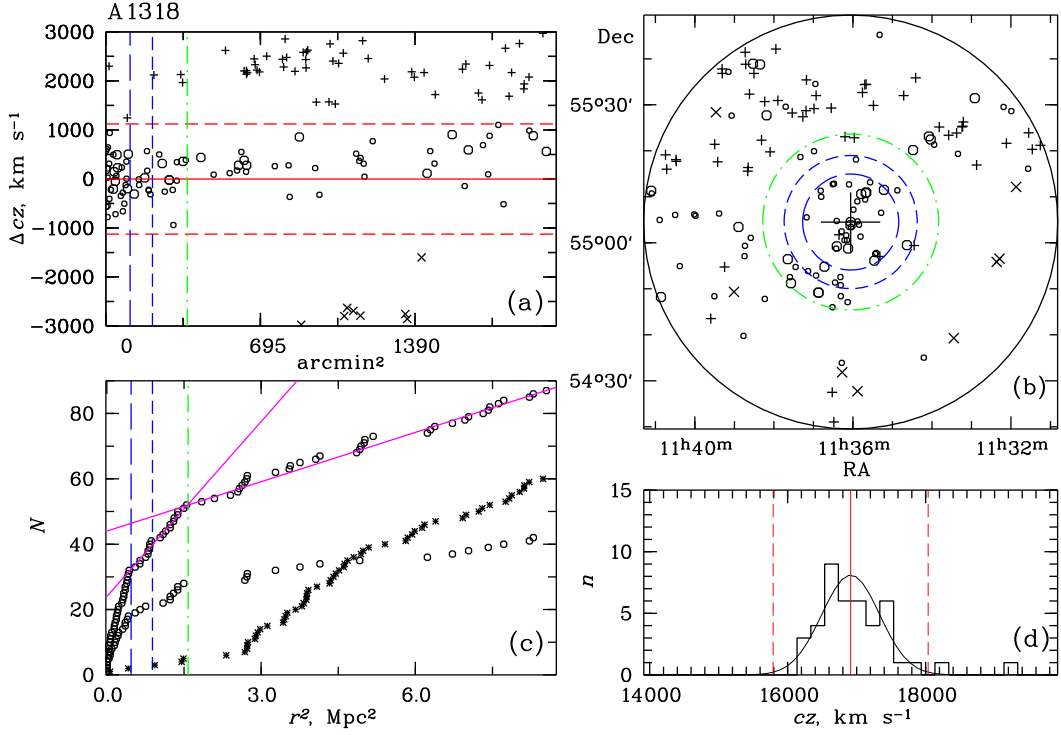}
    \caption{
	Distribution of galaxies in the  A\,1318 cluster. Panel~(a) shows the deviation of line-of-sight velocities
of galaxies from the mean line-of-sight velocity of the cluster inferred from the galaxies located within the $R_{200}$
radius. The  red horizontal dashed lines correspond to  $\pm2.7\sigma$ deviations. The vertical lines show the $R_{200}$ (short dashes),
$R_с$ (long dashes), and $R_{\rm{sp}}$ (dashed-and-dotted line) radii. Larger circles, plus symbols, and crosses indicate
galaxies brighter than \mbox{$M_K^*+1 = -24^{\rm m}$}, field galaxies, respectively.
Panel~(b) shows the sky distribution of galaxies from panel~(a) in the equatorial system (the same designations are used).
The circles outline regions with the radii $R_{200}$ (short dashes), $R_c$ (long dashes), and $R_{\rm{sp}}$ (dashed-and-dotted).
The region under study is limited to the circle of radius 3.5$R_{200}$ (the black solid line). The large cross
indicates the cluster center. Panel~(c) shows the integrated distributions of the total number of galaxies (the top curve)
and early-type galaxies brighter than $M_K<-21\fm5$ (the bottom curve) as a function of the squared distance from the center of the
group. Circles correspond to galaxies shown by circles in panel~(a) and asterisks, to foreground and background galaxies.
The solid magenta lines characterize the behavior of the distribution of galaxies inside and beyond the  $R_{\rm{sp}}$ radius.
Panel~(d) shows the distribution of line-of-sight velocities of all galaxies inside the $R_{200}$ radius
(the solid line for cluster members shows the Gaussian corresponding to the dispersion of line-of-sight velocities).
The solid vertical red line indicates the average line-of-sight  velocity of the cluster and the dashed red lines correspond
to $\pm2.7\sigma$ deviations.}
    \label{clusA1318}
\end{figure*}

We first estimate the average line-of-sight velocity of the cluster, $cz$, and its dispersion $\sigma$, which we then use
to estimate the $R_{200}$ radius. We then determine the number of galaxies inside this radius and again compute the average line-of-sight
velocity of the cluster, $cz$, its dispersion $\sigma$, and infer from it the radius $R_{200}$, etc.
We thus move from the cluster center to iteratively determine the dispersion of line-of-sight velocities of galaxies and
other cluster parameters within the given radius. We considered galaxies with relative velocities greater than $2.7\,\sigma$
to be field objects (e.g., Mamon et al. 2004), selection  criteria usually vary from $2.5\,\sigma$ to
$3.0\,\sigma$.

\begin{figure*}[]
    \setcaptionmargin{5mm} \onelinecaptionstrue \captionstyle{normal}
    \includegraphics[scale=0.53, angle=-90]{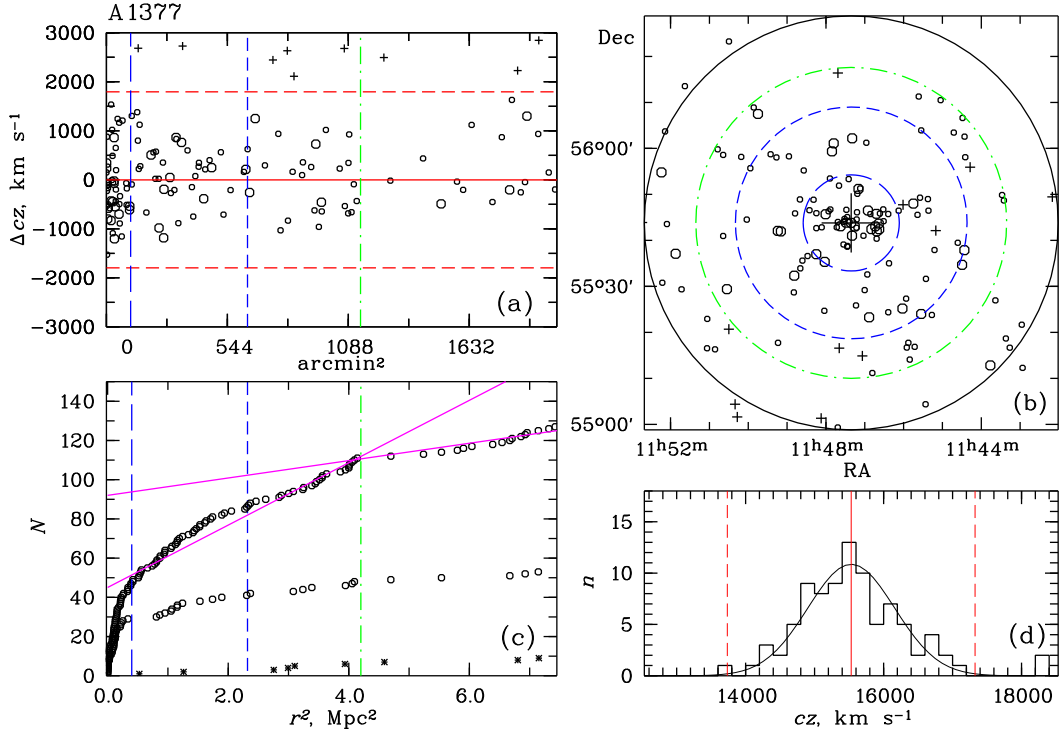}
    \captionstyle{normal}
    \caption{Distribution of galaxies in the  A\,1377 cluster. Designations are the same as in Fig.~1.
    }
    \label{clusA1377}
\end{figure*}

\begin{figure*}[]
    \setcaptionmargin{5mm} \onelinecaptionstrue \captionstyle{normal}
    \includegraphics[scale=0.53, angle=-90]{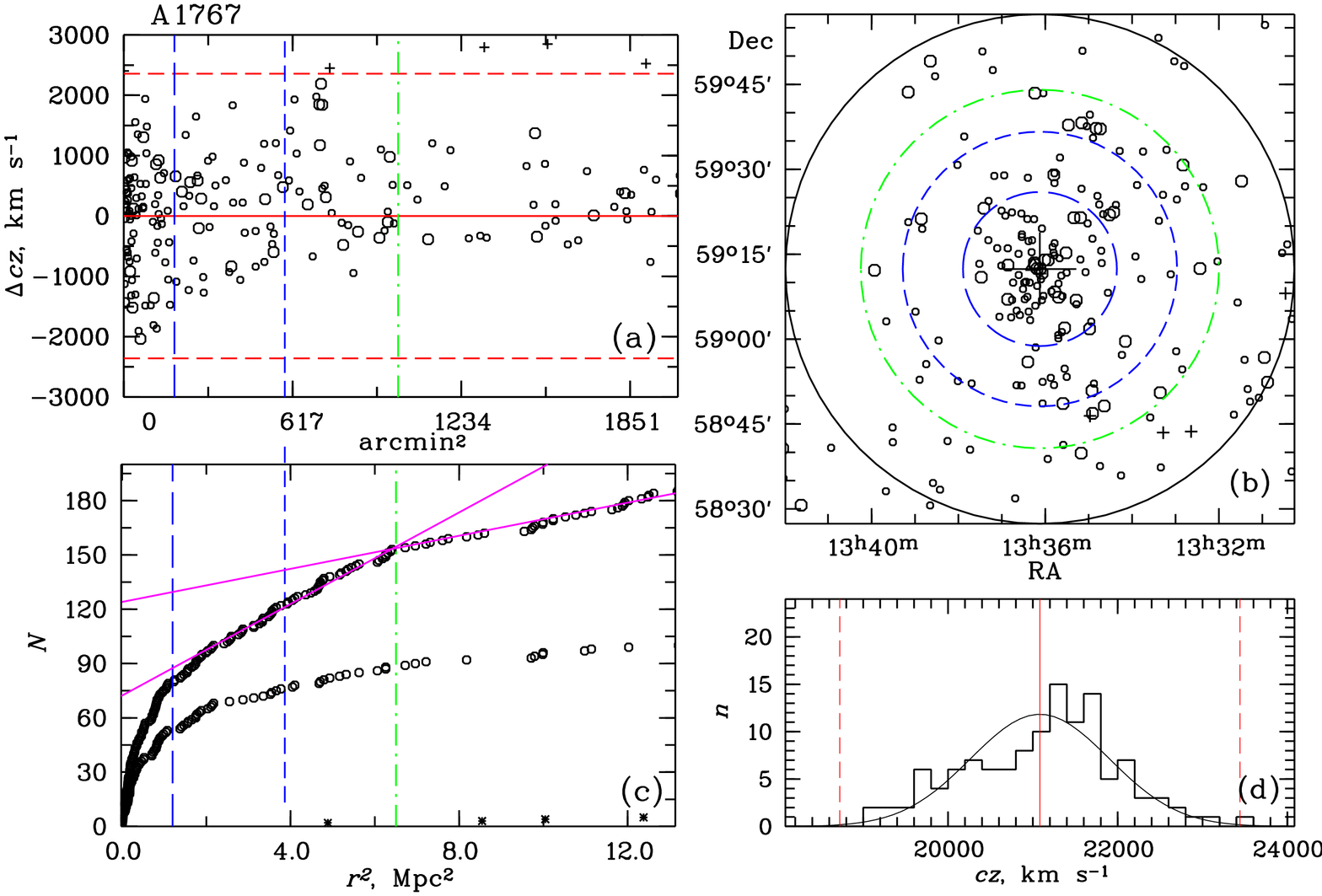}
    \captionstyle{normal}
    \caption{Distribution of galaxies in the  A\,1767 cluster. Designations are the same as in Fig.~1.
    }
    \label{clusA1767}
\end{figure*}

\begin{figure*}[]
    \setcaptionmargin{5mm} \onelinecaptionstrue \captionstyle{normal}
    \includegraphics[scale=0.53, angle=-90]{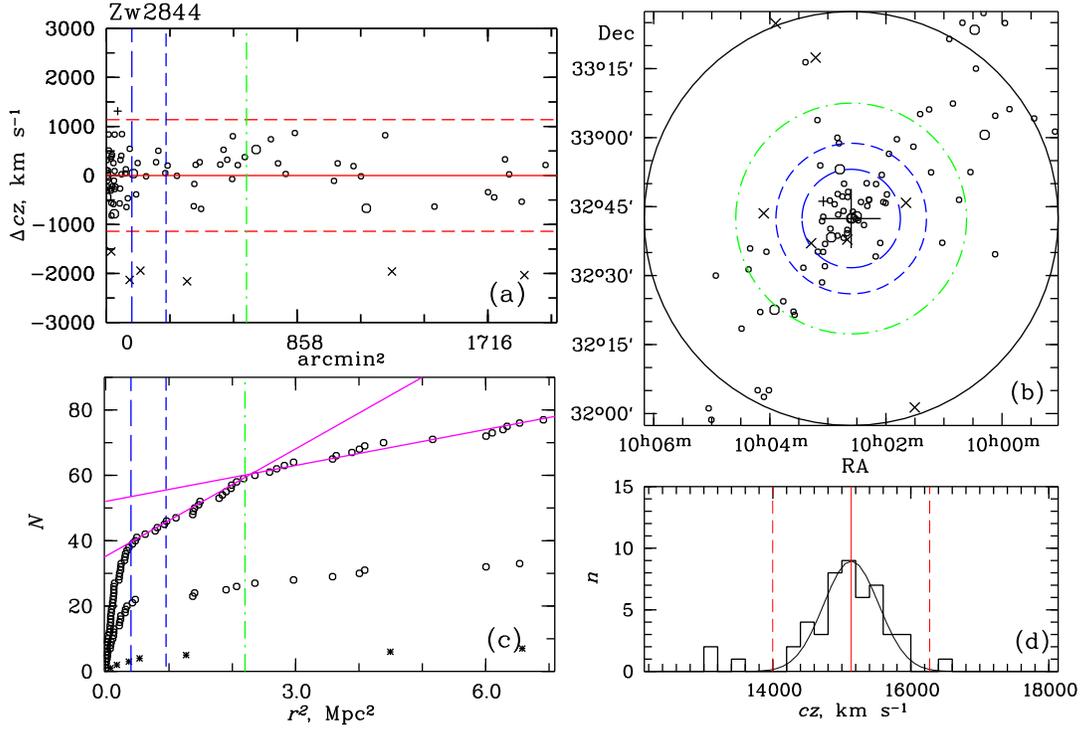}
    \captionstyle{normal}
    \caption{Distribution of galaxies in the  Zw\,2844 cluster. Designations are the same as in Fig.~1.
    }
    \label{clusZw2844}
\end{figure*}

To determine the radius $R_{\rm{sp}}$, it is important to select the closest neighborhoods of groups/clusters of galaxies.
To this end, we use a set of figures that characterize in detail the structure and kinematics of clusters of galaxies, namely:
\begin{list}{}{
\setlength\leftmargin{8mm} \setlength\topsep{1mm}
\setlength\parsep{0mm} \setlength\itemsep{1pt} }
\item [(a)]
 deviation of the line-of-sight velocities of cluster member galaxies and galaxies classified as field objects from the mean
line-of-sight velocity of the cluster or group as a function of the squared radius (clustercentric distance);% (панель~(a));
\item [(b)]
sky positions of galaxies in equatorial coordinates;%(панель~(c));
\item [(c)]
integrated distribution of the number  of all galaxies as a function of the squared radius;%(панель~(b));
\item [(d)]
histogram of the distribution of line-of-sight velocities of all galaxies inside the $R_{200}$ radius.% (панель~(d)).
\end{list}
As an example, we show in Fig.~\ref{clusA1318}--\ref{clusZw2844} such plots for  A\,1318, A\,1377, A\,1767\, and Zw\,2844
clusters, respectively.

\begin{figure*}%[hbt!!]
    \setcaptionmargin{5mm} \onelinecaptionstrue \captionstyle{normal}
    \includegraphics[scale=0.59, bb= 25 110 799 500,clip]{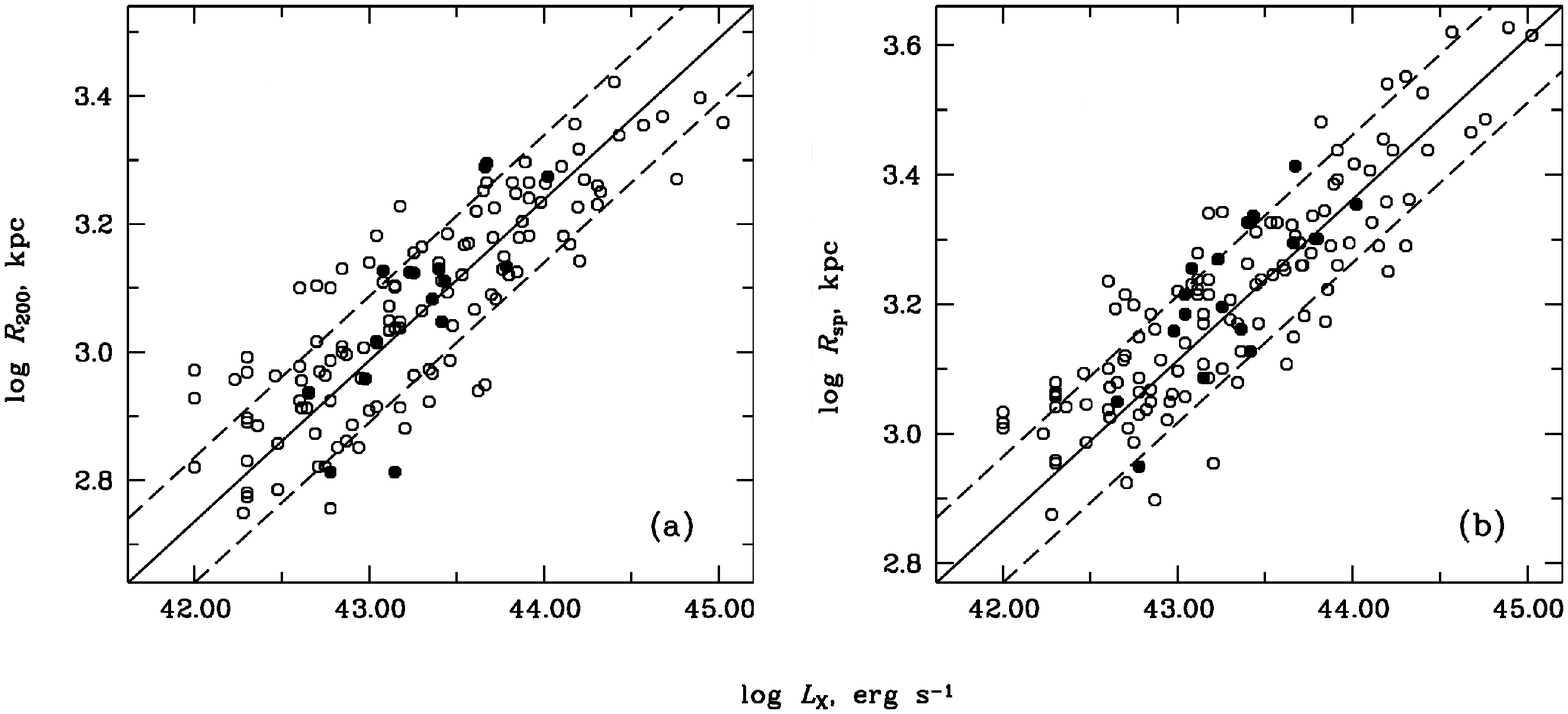}
    \captionstyle{normal}
    \caption{Dependence of radius $R_{200}$ (a) and radius $R_{\rm{sp}}$ (b)
        on the  \mbox{0.1--2.4~keV} X-ray luminosity.
        The solid lines correspond to regression relations
        $R_{\rm{sp}} \propto L_X^{0.24\pm0.03}$ and
        $R_{200} \propto L_X^{0.25\pm0.04}$. The dashed lines show the  $1\sigma$-deviations from these relations.
        The filled circles show groups and clusters with bimodal distribution of line-of-sight velocities.
    }
    \label{RLX}
\end{figure*}

Of special interest is the the projected cluster profile, i.e., the integrated distribution of the number of galaxies as a function
of the squared clustercentric distance as shown in panels~(c) of the analyzed plots (see Figs.~\ref{clusA1318}--\ref{clusZw2844}).
This distribution allows us to visually identify the dense core of the group/cluster, its more tenuous envelope, and the outer region,
where the distribution becomes linear (shown by the solid magenta line in panels~(c) of Figs.~\ref{clusA1318}--\ref{clusZw2844})
in the adopted coordinates, i.e., the distribution of galaxies appears, on the average, uniform
(Kopylov and Kopylova 2015). The dashed and dashed-and-dotted lines in panels~(b) of Figs.~\ref{clusA1318}--\ref{clusZw2844}
show the radius of the virialized region $R_{200}$ and the radius $R_{\rm{sp}}$ beyond which the distribution of the number
of cluster members becomes linear, respectively. We also outlined with the long-dashed curve
the central part of the cluster with radius $R_c$ where the main
sharp increase of the number of galaxies is observed. Panels~(c) (Figs.~\ref{clusA1318}--\ref{clusZw2844})
show the distribution of early-type galaxies brighter than $M_K=-21\,.\!\!^{\rm
m}5$, which was used to refine  these radii. Such galaxies are usually located in central virialized regions of groups/clusters
of galaxies. Our inferred splashback radius  $R_{\rm{sp}}$ (Adhikari et al. 2014: Diemer and Kravtsov 2014) is the mean radius of the apocenters
of the orbits of galaxies that have moved out of the central region of the galaxy cluster. That is, the inferred radius $R_{\rm{sp}}$
separates most of the galaxies that fall for the first time onto the cluster, from the galaxies from the galaxies
that already belong to the cluster. We found for the entire sample \mbox{$\langle R_{\rm{sp}} \rangle =
1.67\pm0.05$}~Mpc with a total spread  0.75--4.24, $\langle R_c
\rangle = 0.78\pm0.03$ with a total spread of 0.30--2.00,
\mbox{$\langle R_{\rm{sp}}/R_{\rm{200с}} \rangle = 1.40\pm0.02$}
or $\langle R_{\rm{sp}}/R_{\rm{200m}} \rangle = 0.88\pm0.02$ (for
\mbox{$4R_{200c} \approx 2.5R_{\rm{200m}}$}), $\langle
R_{\rm{sp}}/R_c \rangle = 2.19\pm0.04$, $\langle R_{200c}/R_c
\rangle= 1.58\pm0.02$. The range of $R_{\rm{sp}}/R_{\rm{200с}}$ ratios for our sample
lies in the \mbox{0.91--1.96} interval and approximately agrees with the
results of simulations, 1.28--2.4 (More et al.2015).

Table~\ref{data1} lists the measured $R_{\rm{sp}}$ and $R_c$ radii and other physical properties of the groups and clusters of
galaxies studied.
%The columns of the table list the following data: (1)---the name of the cluster of galaxies;
%(2)---the heliocentric redshift; (3)---the line-of-sight velocity dispersion with cosmological correction
%\mbox{$(1+z)^{-1}$} applied; (4)---the halo mass $M_{200}$; (5)---the IR-luminosity; (6)---the  \mbox{0.1--2.4}~keV
%X-ray luminosity from X-ray catalogs of galaxy clusters based on  ROSAT observations (see the above papers for more detailed
%references; we recomputed the luminosities using our inferred line-of-sight velocities of clusters and the adopted model).
%(7)---the  $R_{200}$ radius in Mpc; (8)---the $R_{\rm{sp}}$ radius in Mpc; (9)---the $R_с$ radius in Mpc;  (10)---$R_{\rm{sp}}/R_{200}$; (11)---the concentration $\Sigma_5$ of galaxies
%brighter than $M_K=-23\,.\!\!^{\rm m}3$ computed from the distance of the fifth galaxy closest to the center,
%and (12)---the magnitude gap  $\Delta M_{1,4}$ between the brightest and fourth brightest galaxies.

\begin{table*}[]
    \setcaptionmargin{0mm} \onelinecaptionstrue \captionstyle{normal}
    \caption{Examples of relations}
    \label{data2}
    \medskip
   % \small
    \begin{tabular}{r@{--}l|c|r|c}
        \hline
        \multicolumn{2}{c|}{Relation }             & Slope  &  Normalization & Scatter \\
        \hline
        $\log R_{\rm{sp}}$&$\log L_X$               & $0.24\pm0.03$ & $-7.39\pm0.33$ & 0.092 \\
        $\log R_{200}$    &$\log L_X$              & $0.25\pm0.04$ & $-7.60\pm0.34$ & 0.097 \\
        $\log R_c$        &$\log L_X$                  & $0.26\pm0.04$ & $-8.45\pm0.37$ & 0.110 \\
        $\log R_{\rm{sp}}$&$\log(M_{200}/M_{\odot}$) & $0.32\pm0.02$ & $-1.42\pm0.13$ & 0.066 \\
        $\log R_c$        &$\log (M_{200}/M_{\odot}$)    & $0.35\pm0.03$ & $-2.08\pm0.17$ & 0.086 \\
        $\log R_{\rm{sp}}$&$\log (L_K/L_{\odot}$)     & $0.42\pm0.03$ & $-2.00\pm0.17$ & 0.074 \\
        $\log R_{200}$    &$\log (L_K/L_{\odot}$)    & $0.41\pm0.02$ & $-2.10\pm0.12$ & 0.052 \\
        $\log R_c$        &$\log (L_K/L_{\odot}$)        & $0.44\pm0.03$ & $-2.66\pm0.19$ & 0.088 \\
        $\log R_{\rm{sp}}$&$\log R_{200}$           & $1.00\pm0.04$ & $ 0.17\pm0.11$ & 0.064 \\
        $\log R_{\rm{sp}}$&$\log R_c$               & $0.95\pm0.05$ & $ 0.46\pm0.14$ & 0.088 \\
        $\log R_{\rm{sp}}$&$\log \Sigma_5$          & $0.31\pm0.06$ & $ 2.71\pm0.17$ & 0.158 \\
        $\log R_{\rm{sp}}$&$\Delta M_{1,4}$          & $-0.08\pm0.02$& $3.31\pm0.03$& 0.147 \\
        \hline
    \end{tabular}
\end{table*}

\section{RESULTS}

\subsection{Dependences of radius $R_{\rm{sp}}$  on  $L_X$ and $M_{200}$}

By definition $R_{\rm{sp}} > R_{200}$ except for the cases where overestimated dispersion of line-of-sight velocities
is observed in the cluster inside  $R_{200}$---this value can (like in our case) affect the inferred
$R_{200}$ radius. Such clusters of galaxies  (e.g., A\,1825) usually show non-Gaussian distribution of line-of-sight velocities
inside $R_{200}$.

We studied how the $R_{\rm{sp}}$ and $R_c$ radii vary as a function of the properties of groups/clusters of galaxies.
Fig.~\ref{RLX} shows the dependence of  $\log R_{\rm{sp}}$ on X-ray luminosity $\log L_X$, and, for comparison,
a similar relation for the  $\log R_{200}$ radius.
The reported relations (straight lines) are averages of direct and inverse regressions with independent variables swapped.
The dashed lines show the  1\,$\sigma$ deviations from this regression.
Note that the mean-square deviation for the dependence of radius $R_{\rm{sp}}$ is somewhat smaller than for radius $R_{200}$.
The filled circles show merging groups/clusters of galaxies with bimodal distribution of line-of-sight velocities
inside $R_{200}$.
Note that the positions of these structures on common relations do not differ from those of  groups/clusters of galaxies
with Gaussian distribution of line-of-sight velocities.
Our sample also contains the second group of galaxies with no measured X-ray luminosities. These are usually groups of galaxies with
$\sigma \leq$ 400~km~s$^{-1}$.
Table~\ref{data2} lists the parameters of our derived relations---their slopes, intercepts, and scatters.
The splashback radii of groups/clusters that we inferred from observed profiles yield the following relations: $$R_{\rm{sp}} \propto L_X^{0.24\pm0.03},
R_{\rm{sp}} \propto (M_{200}/M_{\odot})^{0.32\pm0.02},$$ \vspace{-3mm}$$R_{\rm{sp}} \propto (L_K/L_{\odot})^{0.42\pm0.03}.$$
The dependence on $L_K$ luminosity  (which characterizes galaxies) is almost twice steeper than the dependence on   $L_X$
luminosity (which characterizes gas).
The dependences of the radius of the radius  $\log R_с$ of the central part of the cluster are, on the average,
steeper by $5\%$ than the corresponding dependences for the $\log R_{\rm{sp}}$ radius.
The smallest scatter is observed for the dependence  $\log R_{\rm{sp}}$ on $\log (M_{200}/M_{\odot}$).
We do not give the dependence between  $\log R_{200}$ and $\log M_{200}/M_{\odot}$ in Table~\ref{data2} because
the two variables are dependent.

\subsection{Dependences of radius $R_{\rm{sp}}$ on $\Sigma_5$ and $\Delta M_{1,4}$}

\begin{figure*}[]
    \setcaptionmargin{5mm} \onelinecaptionstrue \captionstyle{normal}
    \includegraphics[scale=0.59, bb= 25 120 799 500,clip]{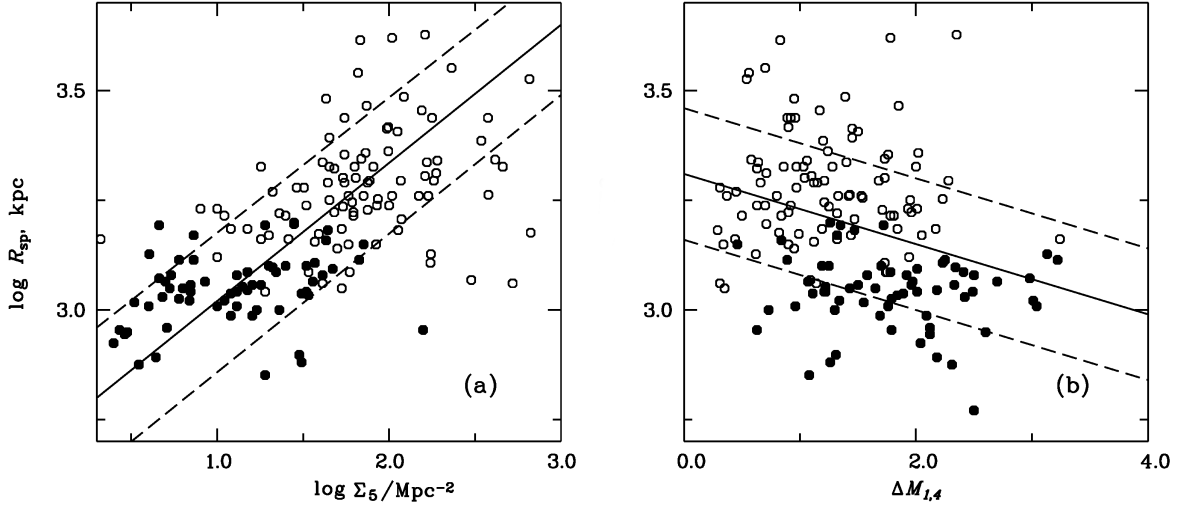}
    \captionstyle{normal}
    \caption{Dependence of radius $R_{\rm{sp}}$ (a)
        on galaxy concentration $\Sigma_5$,
        (b) on magnitude gap $\Delta M_{1,4}$ between the first and fourth brightest galaxies within $0.5\,R_{200}$.
        The filled circles show groups of galaxies with $\sigma \leq 400$\,km\,s$^{-1}$.
        The solid lines show the regression relations
        $R_{\rm{sp}} \propto \Sigma_5^{0.31}$ and
        $R_{\rm{sp}} \propto 10^{-0.08\Delta M_{1,4}}$. The dashed lines show the $1\,\sigma$ deviations from these relations.
    }
    \label{RM14}
\end{figure*}

\begin{figure*}[]
    \setcaptionmargin{5mm} \onelinecaptionstrue \captionstyle{normal}
    \includegraphics[scale=0.59, bb= 25 120 799 500,clip]{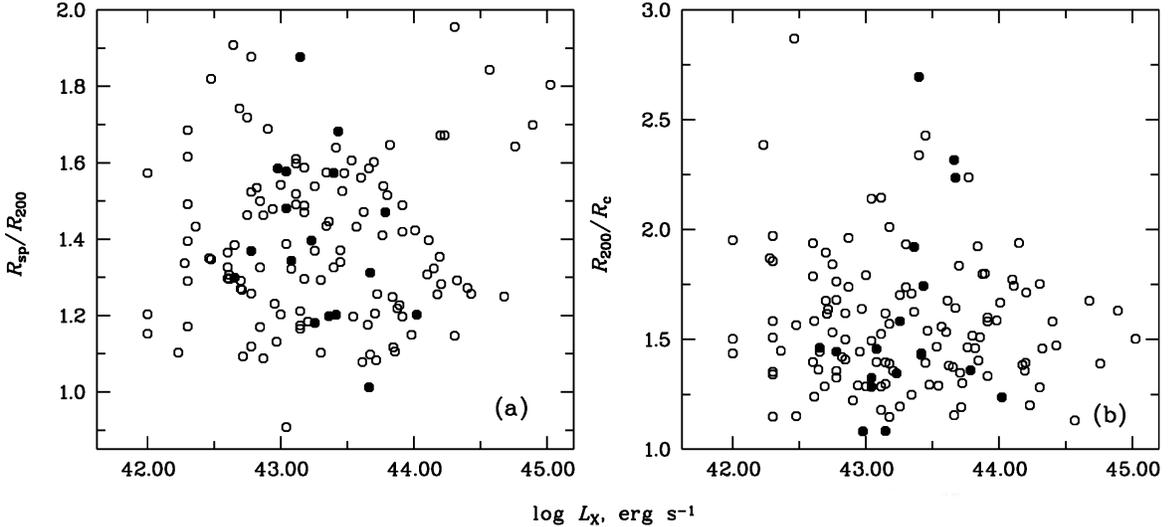}
    \captionstyle{normal}
    \caption{Dependence of the $R_{\rm{sp}}/R_{200}$ (a) and $R_{200}/R_{с}$ (b) ratios
    on the 0.1--2.4~keV X-ray luminosity.
        The filled circles show groups and clusters of galaxies with bimodal distribution
         of line-of-sight velocities.
    }
    \label{RSR2}
\end{figure*}

We measured the concentration of galaxies in groups and clusters as the internal density of galaxies, $\Sigma_5$, computed
from the distance to the fifth nearest galaxy brighter than $M_K=-23\,.\!\!^{\rm m}3$.
The measured concentrations for a significant part of groups of galaxies were published by (Kopylova and Kopylov 2017). The above paper
also provides the measured $K$- and $r$-band  magnitude gap $\Delta M_{1,4}$ between the brightest and fourth brightest
galaxies within the 0.5\,$R_{200}$ radius.
In the cases where the   \mbox{$\Delta M_{1,4} \geq 2.5$} condition is fulfilled dynamically  ``old'' groups with
masses \linebreak \mbox{$M_{200} < 1.43\times 10^{14}\,M_{\odot}$} can be found (Kopylova and Kopylov 2017).
Fig.~\ref{RM14} shows the dependences of radius $R_{\rm{sp}}$ on the concentration of galaxies (Fig.~\ref{RM14}a)
and on the $\Delta M_{1,4}$ magnitude gap (Fig.~\ref{RM14}b). The filled circles show groups of galaxies
with \mbox{$\sigma\,\leq$ 400~km\,s$^{-1}$}.
The relation (the straight line) shown in Fig.~\ref{RM14}a is the average of direct and inverse regressions with independent
variable swapped, \mbox{$R_{\rm{sp}} \propto \Sigma_5^{0.31}$}.
The dashed lines show the 1\,$\sigma$ deviations from this relation.
In our earlier paper~(Kopylova and Kopylov 2017) we showed that the concentration of galaxies in groups and clusters
correlates with richness, X-ray luminosity, mass and the $R_{\rm{sp}}$ radius.
An analysis of Fig.~\ref{RM14}b shows that  $\log R_{\rm{sp}}$ depends  slightly on $\Delta M_{1,4}$ separately
for groups and clusters of galaxies. The combined sample yields \mbox{$R_{\rm{sp}} \propto 10^{-0.081\Delta M_{1,4}}$}, i.e.,
the greater $\Delta M_{1,4}$, the smaller is  the  $R_{\rm{sp}}$ radius for groups of galaxies.
We already showed (Kopylova and Kopylov 2017) that in the  $\Delta M_{1,4}$--$\Sigma_5$ diagram clusters of galaxies
are located in  the region of large concentrations of galaxies and small $\Delta M_{1,4}$, i.e., in the region of
dynamically ``young'' structures. Groups of galaxies are located in the region of low concentration of galaxies and
large magnitude gaps $\Delta M_{1,4}$, i.e., in the region of dynamically  ``old'' structures.

\subsection{Dependences of the  $R_{\rm{sp}}/R_{200}$ ratio on $L_X$, $M_{200}$}

In Section~1 we quoted the results of More et al.(2015), which imply that the $R_{\rm{sp}}$ radius depends on
the dark-matter accretion rate onto the cluster: in the case of high accretion rate this radius is close to
the virial radius, i.e., in our case, it is close to $R_{200}$. Fig.~\ref{RSR2} shows the dependences of
the  $R_{\rm{sp}}/R_{200}$ and $R_{200}/R_c$ radius ratios on the X-ray luminosity of groups/clusters of galaxies.

Fig.~\ref{RSR2} shows that our sample of systems of galaxies appears rather compact on the diagram and
that the $R_{\rm{sp}}/R_{200}$ ratio varies approximately from 1 to 2. We found that this ratio ranges from 1.15 to 1.6
for the bulk of clusters with $\log L_X \in [42.5; 44.5]$. Groups and clusters of galaxies with the radius ratios less
than  1.15 (or $R_{\rm{sp}}/R_{\rm{200m}} < 0.72$) (see Fig.~\ref{RSR2}) can be classified as structures with high mass accretion rates
(FA), and those with the radius ratios greater than 1.6 (or $R_{\rm{sp}}/R_{\rm{200m}} >
1.00$), as structures with slow accretion rates (SA).
As is evident from Fig.~\ref{RSR2}, only two groups are located below $R_{\rm{sp}}/R_{\rm{200с}} \sim 1.08$,
i.e., the actual boundary is at about  $1.08$. Our approximate boundaries are somewhat smaller compared to the
the results of simulations performed by More et al. (2015). Our sample of 157~objects contains 19 FA and 22 SA groups and clusters
of galaxies. FA structures are mostly clusters of galaxies with non-Gaussian distribution of line-of-sight velocities,
signs of mergers with other groups and galaxies near the virial radius, e.g.,  A\,1270, A\,1904, A\,1991,
and NGC\,2563. Among SA clusters of galaxies there are rich clusters, such as A\,1656, A\,1795, A\,2142, and A\,2029,
and poor groups, such as NGC\,7237, IC\,2476, MCG-01-29, which accrete matter (groups, galaxies, and gas)
from large clustercentric distances.

\section{CONCLUSIONS}

Groups and clusters of galaxies have no clear boundaries because of the permanent infall of neighboring groups of galaxies, galaxies,
gas, and dark-matter particles. Therefore determination of their boundaries always remains an issue of great importance
despite the use of known methods that involve a comparison of the distribution of galaxies with results of simulations
or application of the virial theorem to the cluster.
In this paper we report the new observed boundaries of the groups and clusters of galaxies inferred from the galaxies, which we
identify with the splashbasck radius $R_{\rm{sp}}$ equal to the radius of the apocenters of most of the accreted galaxies.
We use our sample of 157 galaxies (based on SDSS data) from the local Universe with redshits \mbox{$0.01 < z < 0.10$}
and masses \linebreak \mbox{$2 \times 10^{13} M_{\odot}<  M_{200} <
2.5 \times 10^{15} M_{\odot}$}. We analyzed the observed profiles of these groups and clusters of galaxies
and determined the radii $R_{\rm{sp}}$ and $R_c$ (the radius of the central region) and studied dependences of these radii
on other properties of groups and clusters. We  reconstructed the observed (projected) profile of each group of cluster of
galaxies---the integrated distribution of the number of all galaxies and early-type galaxies as a function of the squared
distance from the center. This profile shows a steep increase of the number of galaxies, which then becomes linear.
We used the transition point to find $R_{\rm{sp}} > R_{200}$. We used the  steepest part of this profile to estimate the
radius  $R_c$ of the central region of each structure. We studied how  $R_{\rm{sp}}$ and $R_c$ depend
on such properties of groups and clusters of galaxies as their X-ray luminosity, dynamical mass $M_{200}$
inferred from $\sigma$, \mbox{$K$-band} luminosity ($M_K < -21^{\rm m}$), concentration of galaxies $\Sigma_5$,
and the degree to which the brightest galaxy is distinguished, $\Delta M_{1,4}$.

We obtained the following results:

1. The boundary of the dark halo of groups and clusters of galaxies (the $R_{\rm{sp}}$ radius) determined
from galaxies is proportional to the radius  $R_{200}$ of the virialized part estimated from the dispersion of
line-of-sight velocities and to the radius  $R_c$ of the central virialized part.
We found that $R_{\rm{sp}} \propto R_{200}$ and
\mbox{$R_{\rm{sp}} \propto R_c^{0.95\pm0.05}$}.

2. All our measured radii correlate with X-ray luminosity and the dependences have similar slopes. We found the smallest
scatter for the dependence of the  splashback radius on X-ray luminosity:
$R_{\rm{sp}} \propto L_X^{0.24\pm0.03}$ \mbox{($\rm {rms}\!=\!0.092$)}.
We also found that $R_c \propto L_X^{0.26\pm0.04}$ ($\rm {rms}\!=\!0.110$) and
$R_{200} \propto  L_X^{0.25\pm0.03}$ ($\rm {rms} = 0.097$).
The dependences of the  splashback radius on mass $M_{200}$ and  $L_{K,200}$ have even smaller scatter.

3. Note that the $R_{\rm{sp}}/R_{200}$ ratio (Fig.~\ref{RSR2})
for most of the groups and clusters of galaxies varies within a narrow interval from 1 to 2.
However, according to simulations  {More et al. 2015), types of objects with high and low rates of accretion from the
surrounding space can be identified.
We classified 19 groups and clusters as FA structures ($R_{\rm{sp}}/R_{200} \leq 1.15$), and measured
$\langle R_{\rm{sp}}\rangle = 1.42\pm0.10$.
We also classified 22 groups and clusters of galaxies as SA structures   ($R_{\rm{sp}}/R_{200} \geq 1.60$)  with
$\langle R_{\rm{sp}} \rangle = 2.22\pm0.23$.

4. We found a weak dependence of the radius $R_{\rm{sp}}$ of groups and clusters
of galaxies on the  magnitude gap between the first and fourth brightest galaxies, $\Delta M_{1,4}$,
inside the 0.5\,$R_{200}$ radius.

\begin{acknowledgments}
This research has made use of the NASA/IPAC Extragalactic Database
(NED, \url{http://nedwww.}\linebreak\url{.ipac.caltech.edu}),
which is operated by the Jet Propulsion Laboratory, California Institute of
Technology, under contract with the National Aeronautics and Space
Administration, Sloan Digital Sky Survey (SDSS, \url{http://www.sdss.org}),
which is supported by Alfred P. Sloan Foundation, the participant institutes
of the SDSS collaboration, National Science Foundation, and the United
States Department of Energy and Two Micron All Sky Survey (2MASS,
\url{http://www.ipac.}\linebreak\url{.caltech.edu/2mass/releases/allsky/}).
\end{acknowledgments}
 \section*{CONFLICT OF INTEREST}
The authors declare that there is no conflict of interest.

\onecolumngrid
\begin{flushright}
{\it Translated by A.~Dambis}
\end{flushright}


\begin{thebibliography}{99}
%\providecommand{\natexlab}[1]{#1}
%1
\bibitem{Adhikari0}
S.~Adhikari, N.~Dalal, and R.~T.~Chamberlain, J. Cosmology and Astroparticle Physics \textbf{11}, 19 (2014).
%2
\bibitem{Adhikari}
S.~Adhikari, N.~Dalal, and J.~Clampitt, J. Cosmology and Astroparticle Physics \textbf{07}, 22 (2016).
%3
\bibitem{Balogh2}
M.~L.~Balogh, J.~F.~Navarro, and S.~L.~Morris, \apj \textbf{540}, 113  (2000).
%4
\bibitem{Baxter}
E.~Baxter, C.~Chang, B.~Jain, et~al., \apj \textbf{841}, 18 (2017).
%5
\bibitem{Bohringer}
H.~Bo\"{o}hringer, W.~Voges, J.~P.~Huchra et~al., \apjs \textbf{129}, 435  (2000).
%6
\bibitem{Bohringer1}
H.~Bo\"{o}hringer, P.~Schuecker, L.~Guzzo et~al., \aaa~\textbf{425}, 367  (2004).
%7
\bibitem{White}
P.~Busch and S.~D.~M.~White, \mnras \textbf{470}, 4767 (2017).
%8
\bibitem{Carlberg}
R.~G.~Carlberg, H.~K.~C.~Yee, E.~Ellingson, et~al., \apj \textbf{485}, L13 (1997).
%9
\bibitem{Chang}
C.~Chang, E.~Baxter, B.~Jain, et~al., \apj \textbf{864}, 83 (2018).
%10
\bibitem{Contigiani}
O.~Contigiani, H.~Hoekstra, and Y.~M.~Bah\'{e}, \mnras \textbf{485}, 408
  (2019).
%11
\bibitem{Diemer}
B.~Diemer and A.~V.~Kravtsov, \apj \textbf{789}, 1 (2014).
%12
\bibitem{Ebeling}
H.~Ebeling, A.~C.~Edge, H.~Bo\"{o}hringer et~al., \mnras \textbf{301}, 881  (1998).
%13
\bibitem{Ebeling1}
H.~Ebeling, C.~R.~Mullis, R.~B.~Tully, \apj \textbf{580}, 774  (2002).
%14
\bibitem{Fong}
M.~Fong and J.~Han, \mnras \textbf{503}, 4250 (2021).
%15
\bibitem{Gill}
S.~P.~D.~Gill, A.~Knebe, and B.~K.~Gibson, \mnras \textbf{356}, 1327
  (2005).
%16
\bibitem{Gunn}
J.~E.~Gunn and R.~J.~III~Gott \apj \textbf{176}, 1 (1972).
%17
\bibitem{Gott}
R.~J.~III~Gott,  \apj \textbf{186}, 481 (1973).
%18
\bibitem{Haines}
C.~P.~Haines, M.~J.~Pereira, G.~P.~Smith, et~al., \apj \textbf{806}, 101 (2015).
%19
\bibitem{Kopylov07}
A.~I.~Kopylov and F.~G.~Kopylova, \ab~\textbf{62}, 311  (2007).
%20
\bibitem{Kopylova09}
F.~G.~Kopylova and A.~I.~Kopylov, \ab~\textbf{74}, 1  (2009).
%21
\bibitem{Kopylov09}
A.~I.~Kopylov and F.~G.~Kopylova, \ab~\textbf{65}, 205  (2010).
%22
\bibitem{Kopylova11}
F.~G.~Kopylova and A.~I.~Kopylov, Astron. Lett. \textbf{37}, 257  (2011).
%23
\bibitem{Kopylov08}
A.~I.~Kopylov and F.~G.~Kopylova, \ab~\textbf{67}, 17  (2012).
%24
\bibitem{Kopylova13}
F.~G.~Kopylova and A.~I.~Kopylov, Astron. Lett. \textbf{39}, 1  (2013).
%25
\bibitem{Kopylov}
A.~I.~Kopylov and F.~G.~Kopylova, \ab~\textbf{70}, 243  (2015).
%26
\bibitem{Kopylova1}
F.~G.~Kopylova and A.~I.~Kopylov, \ab~\textbf{71}, 257  (2016).
%27
\bibitem{Kopylova10}
F.~G.~Kopylova and A.~I.~Kopylov, \ab~\textbf{72}, 100  (2017).
%28
\bibitem{Kopylova2}
F.~G.~Kopylova and A.~I.~Kopylov, \ab~\textbf{73}, 267  (2018).
%29
\bibitem{Kopylova3}
F.~G.~Kopylova and A.~I.~Kopylov, \ab~\textbf{74}, 365  (2019).
%30
\bibitem{Ledlow}
M.~J.~Ledlow, W.~Voges, F.~N.~Owen, and J.~O.~Burns, \aj \textbf{126}, 2740 (2003).
%31
\bibitem{Mahdavi}
A.~Mahdavi, H.~Bo\"{o}hringer, M.~J.~Geller, and M.~Ramella, \apj \textbf{534}, 114 (2000).
%32
\bibitem{Mamon}
G.~A.~Mamon, T.~Sanchis, E.~Salvador-Sole, and J.~M.~Solanes, \aaa \textbf{414}, 445 (2004).
%33
\bibitem{More1}
S.~More, B.~Diemer, and A.~V.~Kravtsov, \apj \textbf{810}, 36 (2015).
%34
\bibitem{More2}
S.~More, H.~Miyatake, M.~Takada, et~al., \apj \textbf{825}, 39 (2016).
%35
\bibitem{Mulchaey}
J.~S.~Mulchaey, D.~S.~Davis, R.~F.~Mushotsky, and D.~ Burstein, \apjs \textbf{145}, 39 (2003).
%36
\bibitem{Pimbblet}
K.~A.~Pimbblet, \mnras \textbf{411}, 2637 (2010).
%37
\bibitem{Shin}
T.~Shin, S.~Adhikari, E.~J.~Baxter, et~al., \mnras \textbf{487}, 2900
  (2019).
%38
\bibitem{Umetsu}
K.~Umetsu and B.~Diemer, \apj \textbf{836}, 231 (2017).
%39
\bibitem{Zurcher}
D.~Z\'{u}rcher and S.~More, \apj \textbf{874}, 184 (2019).

\end{thebibliography}
\end{document}